\documentclass[prd,nofootinbib,english]{revtex4}

\usepackage{graphicx,float}
\usepackage{amsmath,amssymb,amsfonts}
\usepackage{mathrsfs}
\usepackage{epsfig,color}
\usepackage[thinlines]{easytable}
\usepackage{pdfpages}
\usepackage{array}
\usepackage{cancel}
\usepackage{mathtools}
\usepackage{accents}
\usepackage{subfigure}
\usepackage{enumitem}
\usepackage[dvipsnames]{xcolor}
\usepackage{refstyle}
 \usepackage{hyperref}
\hypersetup{
    colorlinks=true,
    linkcolor=blue,
    filecolor=magenta,
   citecolor=blue
}

\def\be{\begin{equation}}
\def\ee{\end{equation}}
\def\bea{\begin{eqnarray}}
\def\eea{\end{eqnarray}}

\begin{document}
\hfill  USTC-ICTS/PCFT-22-01
\title{Ghost instability in the teleparallel gravity model with parity violations}

\author{Mingzhe Li}
\author{Zhihao Li}
\author{Haomin Rao}
\affiliation{Interdisciplinary Center for Theoretical Study, University of Science and Technology of China, Hefei, Anhui 230026, China}
\affiliation{Peng Huanwu Center for Fundamental Theory, Hefei, Anhui 230026, China}

\begin{abstract}
In this paper we consider the parity violating gravity model within the framework of teleparallel gravity. The parity violations are caused by the couplings of a scalar field to the scalar invariants which are parity-odd and quadratic in the torsion tensor. Totally there are two such type independent invariants, and one of them is the Nieh-Yan density. Through investigations on the cosmological perturbations of this model, we find that in general it suffers from the difficulties of ghost instability in the scalar and vector perturbations. But in the special case only the coupling to the Nieh-Yan density exists, this model is ghost free and reduces to the Nieh-Yan modified Teleparallel Gravity model.
We also analyze the severity of the ghost instability by studying the perturbations around the Minkowski background.
\end{abstract}

\maketitle

\section{introduction}

In recent years, there are many interests in investigating possible parity violations in gravity theories in the literature, partly stimulated by the experimental detections of gravitational waves (GWs) \cite{ligo1,ligo2}
and the developments in the cosmic microwave background radiation (CMB) experiments \cite{CMB1,CMB2}.
A famous and frequently studied parity violating (PV) gravity model is the so-called Chern-Simons (CS) modified gravity \cite{CSgravity1,CSgravity2},
which within the framework of Riemannian geometry modifies general relativity (GR) by a gravitational CS term
$\theta R\tilde{R}$, where $R\tilde{R}\equiv\varepsilon^{\mu\nu\rho\sigma}R_{\mu\nu}^{~~~\alpha\beta}R_{\rho\sigma\alpha\beta}$,
$\theta$ is scalar field,
$R_{\mu\nu\rho\sigma}$ is the Riemann tensor constructed from Levi-Civita connection,
$\varepsilon^{\mu\nu\rho\sigma}=\epsilon^{\mu\nu\rho\sigma}/\sqrt{-g}$ is Levi-Civita tensor,
$\epsilon^{\mu\nu\rho\sigma}$ is antisymmetry symbol and $g$ is determinant of the metric.
The CS modified gravity makes a difference between the amplitudes of the left- and right-handed polarized components of gravitational waves (GWs),
but no difference between their velocities. This is the so-called amplitude birefringence phenomenon.
However, the CS modified gravity suffers from the problem of vacuum
instability because one of the circularly polarized components of GWs becomes a ghost at high frequencies \cite{CSgravity3}, i.e., its kinetic term has a wrong sign.
Further extensions to the CS modified gravity were made in Refs.~\cite{Crisostomi:2017ugk,Gao:2019liu,Zhao:2019xmm}, but these did not stop the ghost mode at high frequencies, as shown explicitly by Eqs. (3.5)-(3.6) and (3.39)-(3.40) in Ref. \cite{Bartolo:2020gsh}. It is very difficult to have a ghost-free PV gravity model within the framework of Riemannian geometry.

To seek the possibilities we may go beyond the Riemannian geometry.
Along this way, the Nieh-Yan modified Teleparallel Gravity (NYTG) model \cite{PVtele1,PVtele2} was recently proposed.
The NYTG model is based on the teleparallel gravity (TG) \cite{Tele,tele2021} which may be considered as a constrained metric-affine theory and is formulated in a spacetime endowed with a metric compatible but curvature free connection, the gravity is identified with the spacetime torsion. One may have a GR equivalent model within the framework of TG (we may call it TGR). The NYTG model \cite{PVtele1,PVtele2}
modifies TGR slightly by the anomalous coupling $\theta \mathcal{T}\widetilde{\mathcal{T}}$ between an axion-like field (it is a pseudo scalar field) $\theta(x)$ and the Nieh-Yan density \cite{Nieh:1981ww}: $\mathcal{T}\widetilde{\mathcal{T}}=(1/2)\varepsilon^{\mu\nu\rho\sigma}\mathcal{T}^{\lambda}_{~\mu\nu}\mathcal{T}_{\lambda\rho\sigma}$ with $\mathcal{T}^{\lambda}_{~\mu\nu}$ being the torsion tensor. The Nieh-Yan density is parity-odd, so at the background with $\partial_{\mu}\theta\neq 0$, the Nieh-Yan coupling term $\theta \mathcal{T}\widetilde{\mathcal{T}}$ violates parity spontaneously.
Also in Refs.~\cite{PVtele1,PVtele2}, the NYTG model has been applied to cosmology.
It was found that this model makes a difference between the propagating velocities of the left- and right-handed polarized components of GWs,
but makes no difference between their amplitudes. This is the so-called velocity birefringence phenomenon. More importantly, through detailed studies on the cosmological perturbations, it was shown in Refs.~\cite{PVtele1,PVtele2} that the NYTG model is ghost-free.
More recently, this model was found to be compatible with the results of most local tests in the Solar System at the post-Newtonian order \cite{Rao:2021azn,Qiao:2021fwi},
the upper limit on its model parameters by the GWs data of LIGO/Virgo Collaboration was obtained in Ref.~\cite{Wu:2021ndf}, and the enhancement of primordial GWs during inflation due to the velocity birefringence of NYTG model and its implications in the air-based GWs experiments were studied in Ref.~\cite{Cai:2021uup}. The application of Nieh-Yan term on the big-bounce cosmology was considered in Ref.~\cite{Bombacigno:2021bpk}.
More recent constraints on the parity violations in gravities can be found in Ref. \cite{Gong:2021jgg}.

Within the TG framework, the Nieh-Yan density $\mathcal{T}\widetilde{\mathcal{T}}$ is not the only parity-odd invariant one can construct from the terms quadratic in the torsion tensor.  A more general model including all the parity-odd quadratic invariants (coupled with the scalar field $\theta(x)$) in the action was considered in Ref.~\cite{PVtele3} and was applied to cosmology.
However, in Ref.~\cite{PVtele3}, only the background evolution of the universe was studied.
In this paper, we will investigate the cosmological perturbations of this general model
\footnote{In Ref.~\cite{PVtele3}, the modifications to TGR by parity-even terms were also considered. We will not consider these further extensions in this paper.}.
We will show that, in general this model is pathological due to the ghost modes appeared in the scalar and vector perturbations, unless all the PV terms except the Nieh-Yan coupling drop out from the action, if so then this model reduces to the NYTG model of Refs. \cite{PVtele1,PVtele2}.
Furthermore, by studying the properties of the ghost modes in Minkowski spacetime,
we will find that it is almost impossible to avoid the ghost instability by lowering the energy scale of the effective field theory.

It deserves mentioning that there is also consideration on PV gravity model \cite{STGPV1} based on the symmetric teleparallel gravity \cite{Nester:1998mp}. However, as it was pointed out in Ref. \cite{ STGPV2}, such kind of models also suffer from the ghost instability in the vector perturbations.

This paper is organized as follows. In Sec.~\ref{Model}
we present the basics of the TG theory and the model which extends NYTG by including extra PV terms in the action.
The cosmological perturbations are studied in Sec.~\ref{perturbations},
we will show with the quadratic actions of the linear perturbations that
in general both the scalar and vector perturbations suffer from the problem of ghost instability,
unless this model reduces to the NYTG model.
To further explore the severity of the ghost instability in this model, we study the perturbations around the Minkowski spacetime in Sec.~\ref{Minkowski}.
The conclusion is presented in Sec.~\ref{conclusion}.

In this paper, we adopt the unit $8\pi G=1$, and use the signature $(+,-,-,-)$ for the metric.
The local space tensor indices are denoted by $A,B,C,...=0, 1, 2, 3$ and by $a, b, c,...=1, 2, 3$ when limiting to spatial components.
They are lowered and raised by the Minkowski metric $\eta_{AB}$ and its inverse $\eta^{AB}$. The spacetime tensor indices are denoted by
 Greek $\mu, \nu, \rho,...=0, 1, 2, 3$ and by Latin $i, j, k,...=1, 2, 3$ when limiting to spatial components. They are lowered and raised by the spacetime metric $g_{\mu\nu}$ and its inverse $g^{\mu\nu}$.
 In addition, we distinguish the spacetime affine connection $\hat{\Gamma}^{\rho}_{~\mu\nu}$
 and its associated covariant derivative $\hat{\nabla}$ from the Levi-Civita connection ${\Gamma}^{\rho}_{~\mu\nu}$
 and its associated covariant derivative ${\nabla}$ respectively.

\section{The Parity Violating gravity model as an extension of NYTG model}\label{Model}

The TG theory can be considered as a constrained metric-affine theory.
It is formulated in a spacetime endowed with a metric $g_{\mu\nu}$ and an affine connection $\hat{\Gamma}^{\rho}_{~\mu\nu}$,
which is curvature free and metric compatible,
\be\label{constrain}
\hat{R}^{\sigma}_{~\rho\mu\nu}=\partial_{\mu}\hat{\Gamma}^{\sigma}_{~\nu\rho}-\partial_{\nu}\hat{\Gamma}^{\sigma}_{~\mu\rho}
+\hat{\Gamma}^{\sigma}_{~\mu\lambda}\hat{\Gamma}^{\lambda}_{~\nu\rho}-\hat{\Gamma}^{\sigma}_{~\nu\lambda}\hat{\Gamma}^{\lambda}_{~\mu\rho}=0~,~
\hat{\nabla}_{\rho}g_{\mu\nu}=\partial_{\rho}g_{\mu\nu}
-\hat{\Gamma}^{\lambda}_{~\rho\mu}g_{\lambda\nu}-\hat{\Gamma}^{\lambda}_{~\rho\nu}g_{\mu\lambda}=0~.
\ee
Without curvature and nonmetricity in the TG theory,
the gravity is identified with spacetime torsion $\mathcal{T}^{\rho}_{~\mu\nu}=2\hat{\Gamma}_{\,[\mu\nu]}^{\rho}$.
In terms of the tetrad $e^{A}_{~\mu}$ and spin connection $\omega^{A}_{~B\mu}$, which relates the metric and affine connection through $g_{\mu\nu}=\eta_{AB}e^{A}_{~\mu}e^{B}_{~\nu}$ and
$\hat{\Gamma}^{\rho}_{~\mu\nu}=e_{A}^{~\,\rho}(\partial_{\mu}e^A_{~\nu}+\omega^A_{~B\mu}e^B_{~\nu})$,
the torsion tensor can be expressed as
\be\label{torsion tensor}
\mathcal{T}^{\rho}_{~\mu\nu}=2e_{A}^{~\,\rho}(\partial_{[\mu}e^{A}_{~\nu]}+\omega^{A}_{~B[\mu}e^{B}_{~\nu]})~.
\ee
The teleparallel constraints (\ref{constrain}) dictate that the spin connection can be in general expressed as,
\be\label{omega}
\omega_{~B \mu}^{A}=(\Lambda^{-1})^{A}_{~C} \partial_{\mu} \Lambda_{~B}^{C}~,
\ee
where $\Lambda^{A}_{~B}$ is arbitrary element of Lorentz transformation matrix which is position dependent and satisfies the relation $\eta_{AB}\Lambda^A_{~C}\Lambda^B_{~D}=\eta_{CD}$ at any spacetime point.
Therefore, the tetrad $e^{A}_{~\mu}$ and the Lorentz matrix $\Lambda^{A}_{~B}$ can be regarded as the basic variables of the TG theory.

The TGR model, as the GR equivalent TG model, has the following action,
\bea\label{TG action}
S_{TGR}=\frac{1}{2}\int d^4x ~{\|e\|}\mathbb{T}\equiv\int d^4x~{\|e\|} \left(-\frac{1}{2}\mathcal{T}_{\mu}\mathcal{T}^{\mu}+\frac{1}{8}\mathcal{T}_{\alpha\beta\mu}\mathcal{T}^{\alpha\beta\mu}
+\frac{1}{4}\mathcal{T}_{\alpha\beta\mu}\mathcal{T}^{\beta\alpha\mu}\right)~,
\eea
where ${ \|e\|}=\sqrt{-g}$ is the determinant of the tetrad, $\mathbb{T}$ is the torsion scalar, and $\mathcal{T}_{\mu}=\mathcal{T}^{\alpha}_{~~\mu\alpha}$ is the torsion vector.
Since we have the identity $-{R}(e)=\mathbb{T}+2{\nabla}_{\mu}\mathcal{T}^{\mu}$,
the action (\ref{TG action}) is identical to the Einstein-Hilbert action up to a surface term,
where the curvature scalar $R(e)$ is defined by the Levi-Civita connection and considered
as being fully constructed from the metric, and in turn from the tetrad.

As mentioned before, the NYTG model \cite{PVtele1,PVtele2} modifies the TGR model by a coupling Lagrangian between an axion-like field and the Nieh-Yan density. Generally we should also consider its own dynamics of the axion-like field and take other matter into account, so the full action of the NYTG model is
\bea\label{NYTG}
S_{NYTG}=\int d^4x~ {\|e\|}\left[\frac{1}{2}\mathbb{T}
+\frac{c}{4}\,\theta\,\mathcal{T}_{\lambda\mu\nu}\widetilde{\mathcal{T}}^{\lambda\mu\nu}+\frac{1}{2}\nabla_{\mu}\theta\nabla^{\mu}\theta-V(\theta)\right]+S_m~,
\eea
where $c$ is the coupling constant
and $\widetilde{\mathcal{T}}^{\lambda\mu\nu}=(1/2)\varepsilon^{\mu\nu\rho\sigma}\mathcal{T}^{\lambda}_{~~\rho\sigma}$.
Other matter with the action $S_m$ is assumed to be coupled to spacetime minimally through the tetrad. At the background in which the axion-like field has non-zero spacetime derivatives, the Nieh-Yan coupling term breaks parity spontaneously.
Because only the first-order derivatives of the basic variables appears in the action,
the NYTG model can avoid the Ostrogradski ghost mode originated from higher-order derivatives.
Detailed analyses on the cosmological perturbations \cite{PVtele1,PVtele2} showed that this model is ghost free and exhibits a velocity birefringence phenomenon of GWs, which is an explicit signal of parity violation.

Besides the Nieh-Yan density, there are other scalar invariants which are parity-odd and quadratic in the torsion tensor within the TG framework.
Totally we have the following four invariants of this type:
\be
\mathcal{P}_{1}=\frac{1}{2}\varepsilon^{\mu\nu\rho\sigma}\mathcal{T}^{\lambda}_{~\mu\nu}\mathcal{T}_{\lambda\rho\sigma}~,~
\mathcal{P}_{2}=\varepsilon^{\mu\nu\rho\sigma}\mathcal{T}_{\mu}\mathcal{T}_{\nu\rho\sigma}~,~
\mathcal{P}_{3}=\varepsilon^{\mu\nu\rho\sigma}\mathcal{T}^{\lambda}_{~\mu\nu}\mathcal{T}_{\rho\sigma\lambda}~,~
\mathcal{P}_{4}=\varepsilon^{\mu\nu\rho\sigma}\mathcal{T}_{\mu\nu}^{~~\lambda}\mathcal{T}_{\rho\sigma\lambda}~,
\ee
where $\mathcal{P}_{1}$ is just the Nieh-Yan density. None of these terms hide higher derivatives.
 However, these invariants are not fully independent. There exist two identities: $\mathcal{P}_{3}=-\mathcal{P}_{1}+\mathcal{P}_{2}$ and $\mathcal{P}_{4}=(1/2)\mathcal{P}_{1}-\mathcal{P}_{2}$,
so only two of them are independent, this is consistent with the result of Ref.~\cite{PVtele3}.
For convenience we need only to consider $\mathcal{P}_{1}$ and $\mathcal{P}_{2}$ as independent parity-odd invariants.
As a generalization of the NYTG model, we will in this paper consider the couplings of $\theta(x)$ to both $\mathcal{P}_{1}$ and $\mathcal{P}_{2}$ as in Ref. \cite{PVtele3}.
In another word, the PV gravity model we will study in more detail in this paper has the full action:
\bea\label{model}
S=\int d^4x~ {\|e\|}\left[\frac{1}{2}\mathbb{T}
+\frac{1}{4}\,\theta\left(c_{1}\mathcal{P}_{1}+c_{2}\mathcal{P}_{2}\right)
+\frac{1}{2}\nabla_{\mu}\theta\nabla^{\mu}\theta-V(\theta)\right]+S_m~,
\eea
where both $c_1$ and $c_2$ are coupling constants, if $c_{2}=0$ the model returns to the NYTG model. This action is our starting point in this paper.

Like the NYTG model, only the first-order derivatives of the basic variables appear in the action (\ref{model}),
this general model can avoid the Ostrogradski ghost mode from higher-order derivatives.
But this does not mean this model is ghost free, it only means that if ghost modes appear, they are not the Ostrogradski type.
The equations of motion follow from the variations with respect to $e^{A}_{~\mu}$ and $\Lambda^{A}_{~B}$ separately:
\bea
 G^{\mu\nu}+N^{\mu\nu}&=&T^{\mu\nu}+T^{\mu\nu}_{\theta}~,\label{eom1}\\
 N^{[\mu\nu]}&=&0~,\label{eom2}
\eea
where $G^{\mu\nu}$ is the Einstein tensor, $T^{\mu\nu}=-(2/\sqrt{-g})(\delta S_m/\delta g_{\mu\nu})$
and $T^{\mu\nu}_{\theta}=[V(\theta)-\nabla_{\alpha}\theta\nabla^{\alpha}\theta/2]g^{\mu\nu}+\nabla^{\mu}\theta\nabla^{\nu}\theta$
are the energy-momentum tensors for the matter and the scalar field $\theta$ respectively,
and
\be\label{nmn}
 N^{\mu\nu}=
\frac{c_{2}}{2}\left[\hat{\nabla}^{\mu}(\theta\tilde{\mathcal{T}}^{\nu})-\varepsilon^{\mu\nu\rho\sigma}\hat{\nabla}_{\rho}(\theta\mathcal{T}_{\sigma})
-g^{\mu\nu}\tilde{\mathcal{T}}^{\rho}\partial_{\rho}\theta
-\theta\left(2\tilde{\mathcal{T}}^{(\mu\nu)\rho}\mathcal{T}_{\rho}+\frac{1}{2}g^{\mu\nu}\mathcal{P}_{1}
-\mathcal{T}^{\mu}\tilde{\mathcal{T}}^{\nu}\right)
\right]+c_{1}\tilde{\mathcal{T}}^{\mu\nu\rho}\partial_{\rho}\theta~,
\ee
where $\tilde{\mathcal{T}}^{\mu}=g_{\rho\nu}\tilde{\mathcal{T}}^{\rho\mu\nu}$.
Similar to most modified TG models, the equation of motion (\ref{eom2}) from the variation of $\Lambda^{A}_{~B}$ is
not independent of Eq. (\ref{eom1}), it is just the antisymmetric part of the latter.
As explained in Ref.~\cite{PVtele2}, this is due to the local Lorentz invariance of the action,
any change caused by $\delta\Lambda^{A}_{~B}$ can always be equivalent to the change caused by $\delta e^{A}_{~\mu}$,
so requiring the action to take the extremum under $\delta e^{A}_{~\mu}$
already includes the case where the action takes the extremum under $\delta\Lambda^{A}_{~B}$.
There is another equation following from the variation of the action (\ref{model}) with respect to $\theta$,
\be\label{eom3}
{\square}\theta+V^{(1)}-\frac{c_{1}}{4}\mathcal{P}_{1}-\frac{c_{2}}{4}\mathcal{P}_{2}=0~,
\ee
where ${\square}=g^{\mu\nu}{\nabla}_{\mu}{\nabla}_{\nu}$ and $V_{\theta}$ is the first derivative of the potential to the scalar field $\theta$.
All these equations of motion are consistent with the Bianchi identity $\nabla_{\mu}G^{\mu\nu}=0$
and the covariant conservation law $\nabla_{\mu} T^{\mu \nu}=0$.

Again, like most modified TG models,
this model (\ref{model}) has two kinds of gauge symmetries:
the diffeomorphism invariance and the local Lorentz invariance, the local Lorentz transformation makes the following change:
\be\label{LT}
e^{A}_{~\mu}\rightarrow(L^{-1})^{A}_{~B}e^{B}_{~\mu}~,~ \Lambda^{A}_{~B}\rightarrow\Lambda^{A}_{~C}L^{C}_{~B}~,
\ee
where $L^{A}_{~B}$ is also the element of Lorentz matrix.
It's easy to prove that the metric $g_{\mu\nu}$ and the torsion tensor $\mathcal{T}_{~\mu \nu}^{\rho}$
are invariant under the local Lorentz transformation (\ref{LT}), so is the action (\ref{model}).
Due to the local Lorentz invariance, we can always choose the gauge $\Lambda^{A}_{~B}=\delta^{A}_{~B}$, i.e., $\omega^{A}_{~B\mu}=0$.
This is the Weitzenb\"{o}ck connection which had been frequently adopted in the literature.
This gauge is also called the Weitzenb\"{o}ck gauge.
Note that once the Weitzenb\"{o}ck gauge $\omega^{A}_{~B\mu}=0$ is adopted,
the model will no longer be local Lorentz invariant.

Even taking the Weitzenb\"{o}ck gauge, the model still has sixteen basic variables in the gravity sector, while GR has only ten.
At the same time, the model also has sixteen gravitational field equations implied in Eqs. (\ref{eom1}) and (\ref{eom2}).
One can find from the expression (\ref{nmn}) that the extra six equations in (\ref{eom2}) are second order in time derivatives unless $c_{2}=0$.
So naively, it can be expected that the general model (\ref{model}) presents six more dynamical degrees of freedom than the case of GR with a minimally coupled scalar field and other matter.
We will confirm this judgement in Sec.~\ref{perturbations}.
It is some of these extra degrees of freedom bring ghost instabilities.

\section{Cosmological perturbations and quadratic actions}\label{perturbations}

From now on, we apply the model (\ref{model}) to cosmology.
This model corresponds to a special case of a more general model considered in Ref.~\cite{PVtele3} where
the dynamics of the cosmological background has been studied in detail.
Hence, we will not consider the evolution of the background any more.
We will focus on the linear cosmological perturbations around the Friedmann-Robertson-Walker (FRW) background.
For simplicity, we take the Weitzenb\"{o}ck gauge $\omega^{A}_{~B\mu}=0$, and we only consider the spatially-flat FRW universe as the background because this is enough for us to see the difficulties of this model.

We use the following parametrization for perturbed tetrad \cite{PVtele2, tetradper1}:
\bea\label{tetradperturbation}
& & e^{0}_{\ 0}=a(1+A)~,~\nonumber e^{0}_{\ i}=a(\beta_{,i}+\beta_{i}^{V})~,
~\nonumber e^{c}_{\ 0}=a\delta_{ci}(\chi_{,i}+\chi_{i}^{V})~,\nonumber\\
& & e^{c}_{\ i}=a\delta_{cj}[ (1-\psi)\delta_{ij}+\alpha_{,ij}+\alpha_{j,i}^{V}-
              \epsilon_{ijk}(\lambda_{,k}+\lambda_{k}^{V})+\frac{1}{2}h^{T}_{ij}]~,
\eea
where $a$ is the scale factor of the universe and the subscript $``,i"$ means $\partial_{i}$.
So the perturbed metric components have the familiar forms:
\bea
& &g_{00}=a^{2}(1+2A)~,~ g_{0i}=-a^{2}(B_{,i}+B_{i}^{V})~,\nonumber\\
& &g_{ij}=-a^{2}[(1-2\psi)\delta_{ij}+2\alpha_{,ij}+\alpha_{i,j}^{V}+\alpha_{j,i}^{V}+h^{T}_{ij}]~,
\eea
where $B=\chi-\beta$ and $B^{V}_{i}=\chi^{V}_{i}-\beta^{V}_{i}$. Besides the familiar scalar perturbations ($A, B, \psi, \alpha$), vector perturbations ($B_i^V, \alpha^V_i$), and tensor perturbations $h^T_{ij}$ in the metric, the parametrization of tetrad brings six extra variables, which are scalar perturbation $\lambda, \chi+\beta$ and vector perturbation $\lambda_i^V, \chi_i^V+\beta_i^V$. All the vector perturbations are transverse and denoted by the superscript $V$, both the tensor perturbations are transverse and traceless and denoted by the superscript $T$.
In addition, the scalar field $\theta$ is decomposed as $\theta(\eta, \vec{x})=\bar{\theta}(\eta)+\delta\theta(\eta, \vec{x})$ with $\eta$ being the conformal time. In addition, we should also consider the perturbations to the matter contained in $S_m$.

With this scheme one can obtain the perturbation equations directly for the model (\ref{model}), but in order to see whether there are dangerous propagating modes it is better to consider the quadratic actions for the linear perturbations. When applying the model to the early universe, such as the inflationary epoch, in that case the scalar field $\theta$ may be considered as the inflaton, we need to quantize these perturbations to have a mechanism for generating the primordial perturbations which seed the large scale structure at later time. For this purpose, the quadratic actions are indispensable.
For simplicity in the following discussions, we ignore other matter so that $S_m=0$ in Eq. (\ref{model}) from now on.

Even we have taken the Weitzenb\"{o}ck connection, the diffeomorphism invariance is preserved, it is safe to take the unitary gauge $\delta\theta=0,~\alpha=0,~\alpha_{i}^{V}=0$.
This simplifies the calculations, for example, the gauge invariant scalar perturbation $\zeta=-(\psi+\mathcal{H}\delta\theta/\theta')$ representing the curvature perturbation of the hypersurfaces of constant $\theta$ field reduces to $-\psi$ under the unitary gauge, where $\mathcal{H}=a'/a$ is the conformal Hubble rate. With these we may choose $A$, $\zeta$, $B$, $\beta$, $\lambda$, $B_{i}^{V}$, $\beta_{i}^{V}$, $\lambda_{i}^{V}$ and $h^{T}_{ij}$ as independent variables.

As mentioned before,  there is no higher-order derivative in the action of the extended model (\ref{model}), so it avoids the Ostrogradski ghost mode.
But the couplings $\theta\left(c_{1}\mathcal{P}_{1}+c_{2}\mathcal{P}_{2}\right)$ are possible to cause mixings among perturbation variables, such as $x'y'=(1/2)[(x'+y')^2-(x'-y')^2]\equiv(1/2)(z_1'^2-z_2'^2)$, here the prime represents the derivative with respect to the conformal time. Such mixings have the potential to introduce ghost modes into the quadratic actions. We will investigate the quadratic actions for the scalar, vector, and tensor perturbations of the model (\ref{model}) separately in the following subsections.

\subsection{Quadratic action for scalar perturbations}\label{actionscalar}

For the model (\ref{model}) with $S_m=0$, one can obtain directly the quadratic action for the scalar perturbations:
\bea\label{scalar1}
\nonumber S^{(2)}_{S}&=&-\int d\eta d^{3}k~ a^{2} \Big\{3{\zeta'}^{2}-6\mathcal{H}\zeta'A-k^{2}(2A+\zeta)\zeta+a^{2}V(\theta) A^{2}
             +2k^{2}(\zeta'-\mathcal{H}A)B\\
& &-(2c_{1}+3c_{2})\theta'k^{2}\zeta\lambda-c_{2}\theta k^{2}\left[\beta'\lambda'-k^{2}(B+\beta)\lambda
   +(\zeta-A)\lambda'\right]\Big\}~,
\eea
we have changed to the momentum space in terms of Fourier transformations.
We also simply mark $A^{*}B$ as $AB$, $A^{*}A$ as $A^{2}$, and so on.
No matter what the values the coupling constants $c_1$ and $c_2$ have, the variables $A$ and $B$ are non-dynamical fields and the variations of the quadratic action with them lead to the following two constraints:
\bea
& &\label{cscalar1} 2(\mathcal{H}A-\zeta')-c_{2}\theta k^{2}\lambda=0~,\\
& &\label{cscalar2} 6\mathcal{H}\zeta'+2k^{2}\zeta-2a^{2}V(\theta)A+2\mathcal{H}k^{2}B-c_{2}\theta k^{2}\lambda'=0~.
\eea
These constraint equations are just used to solve the non-dynamical variables $A$ and $B$.
One can eliminate these two non-dynamical variables from the action (\ref{scalar1}) by substituting the constraints (\ref{cscalar1}) and (\ref{cscalar2}) back into it. After that,
the quadratic action for scalar perturbations can be expressed as
\bea
& &\nonumber S^{(2)}_{S}=\int d\eta d^{3}k\, \Bigg\{ \frac{a^{2}{\theta'}^{2}}{2\mathcal{H}^{2}}\left({\zeta'}^{2}-k^{2}\zeta^{2}\right)
+2c_{1}a^{2}\theta'k^{2}\zeta\lambda+c_{2}a^{2}\theta k^{2}\bigg[\lambda'\beta'-\frac{1}{\mathcal{H}}\lambda'\zeta'
-\Big(1-\frac{{\theta'}^{2}}{2\mathcal{H}^{2}}\Big)\lambda\zeta'
\\
& & \quad\quad\quad\quad\quad\quad\quad\quad
-k^{2}\lambda\beta
+\Big(\frac{k^{2}}{\mathcal{H}}+2\frac{\theta'}{\theta}-2\mathcal{H}\Big)\lambda\zeta
+\frac{c_{2}(\theta\theta'^{2}+2\mathcal{H}\theta'-2\mathcal{H}^{2}\theta)}{4\mathcal{H}^{2}}k^{2}\lambda^{2}\bigg]
 \Bigg\}~.\label{scalar1.5}
\eea
From this action, one will see that
whether the coefficient $c_2$ vanishes or not determines whether the scalar perturbations of the model (\ref{model}) contain ghost modes.

In the case where $c_2=0$, the model (\ref{model}) reduces to the NYTG model, the quadratic action (\ref{scalar1.5}) reduces to the following one,
\be
S^{(2)}_{S}=\int d\eta d^{3}k\, \left[ \frac{a^{2}{\theta'}^{2}}{2\mathcal{H}^{2}}\left({\zeta'}^{2}-k^{2}\zeta^{2}\right)
+2c_{1}a^{2}\theta'k^{2}\zeta\lambda
\right]~.
\ee
Evidently there is no ghost mode in this action. In fact $\lambda$ in this case is a Lagrange multiplier and leads to $\zeta=0$, all the scalar perturbations are non-dynamical. This is a special feature of the linear perturbation theory of the NYTG model when applied to spatially flat universe \cite{PVtele1}. If the background of the universe has nonzero spatial curvature, there is one dynamical scalar mode which is normal and healthy \cite{PVtele2}.

In the more general case with $c_2\neq0$, the kinetic terms $\lambda'\beta'$ and $\lambda'\zeta'$ led by $c_2$ in the action (\ref{scalar1.5}) generally contain ghost modes, as mentioned in the paragraph just before this subsection.
In order to explicitly see how many dynamical degrees of freedom and how many ghost modes there are,
we define new perturbation variables $\gamma_i$ with $i=1,2,3$ in terms of the old variables $\zeta$, $\beta$, and $\lambda$:
\bea
& &\nonumber \gamma_{1}=\frac{a\sqrt{|3+3\tau_{1}-\tau_{1}^{2}}|}{\sqrt{2}\mathcal{H}\theta'(\tau_{1}-\tau_{2})(\tau_{1}-\tau_{3})}
\left[(1-\tau_{2}-\tau_{3})\theta^{\prime 2}\zeta+\mathcal{H}\theta^{\prime 2}(\tau_{2}+\tau_{3})\beta
-c_{2}\mathcal{H}(2+\tau_{2}\tau_{3})k^{2}\lambda\right]~,\\
& &\nonumber \gamma_{2}=\frac{a\sqrt{|3+3\tau_{2}-\tau_{2}^{2}}|}{\sqrt{2}\mathcal{H}\theta'(\tau_{2}-\tau_{1})(\tau_{2}-\tau_{3})}
\left[(1-\tau_{1}-\tau_{3})\theta^{\prime 2}\zeta+\mathcal{H}\theta^{\prime 2}(\tau_{1}+\tau_{3})\beta
-c_{2}\mathcal{H}(2+\tau_{1}\tau_{3})k^{2}\lambda\right]~,\\
& &          \gamma_{3}=\frac{a\sqrt{|3+3\tau_{3}-\tau_{3}^{2}}|}{\sqrt{2}\mathcal{H}\theta'(\tau_{3}-\tau_{1})(\tau_{3}-\tau_{2})}
\left[(1-\tau_{1}-\tau_{2})\theta^{\prime 2}\zeta+\mathcal{H}\theta^{\prime 2}(\tau_{1}+\tau_{2})\beta
-c_{2}\mathcal{H}(2+\tau_{1}\tau_{2})k^{2}\lambda\right]~,\label{trans}
\eea
where $\tau_{1}$, $\tau_{2}$ and $\tau_{3}$ are the three solutions of the equation $1-2\tau-\tau^{2}+\tau^{3}=0$, and numerically $\tau_{1}\approx 0.445$, $\tau_{2}\approx1.802$ and $\tau_{3}\approx-1.247$.
It can be verified that the transformation (\ref{trans}) is linearly reversible.
Then we can express the action (\ref{scalar1.5}) in terms of the new perturbation variables as,
\be\label{scalar2}
S^{(2)}_{S}=\int d\eta d^{3}k\,
\Big\{{\gamma_{1}'}^{2}+{\gamma_{2}'}^{2}-{\gamma_{3}'}^{2}
+\mathring{\mathcal{C}}_{12}\gamma'_{1}\gamma_{2}+\mathring{\mathcal{C}}_{13}\gamma'_{1}\gamma_{3}+\mathring{\mathcal{C}}_{23}\gamma'_{2}\gamma_{3}
+\hat{\mathcal{C}}_{1}\gamma_{1}^{2}+\hat{\mathcal{C}}_{2}\gamma_{2}^{2}+\hat{\mathcal{C}}_{3}\gamma_{3}^{2}
+\tilde{\mathcal{C}}_{12}\gamma_{1}\gamma_{2}+\tilde{\mathcal{C}}_{13}\gamma_{1}\gamma_{3}+\tilde{\mathcal{C}}_{23}\gamma_{2}\gamma_{3}\Big\}~,
\ee
where
\bea
& &\nonumber \mathring{\mathcal{C}}_{ij}=\frac{2(\tau_{i}-\tau_{j})}{\sqrt{|3+3\tau_{i}-\tau_{i}^{2}||3+3\tau_{j}-\tau_{j}^{2}|}}
             \left(2\frac{\theta''}{\theta'}-\frac{\theta'}{\theta}\right)~,\\
& &\nonumber \hat{\mathcal{C}}_{i}=(i^{2}-3i+1)k^{2}+\frac{1}{|3+3\tau_{i}-\tau_{i}^{2}|}
             \Bigg\{4\tau_{i}\mathcal{H}^{2}+2\left[6+\tau_{i}-3\tau_{i}^{2}+2\frac{c_{1}}{c_{2}}(2-\tau_{i}^{2})\right]\mathcal{H}
             \frac{\theta'}{\theta}
             +\frac{1}{2}\left(\tau_{i}^{2}-3\tau_{i}-5-\frac{2\tau_{i}}{\theta^{2}} \right)\theta^{\prime 2}\\
& &\nonumber \quad\quad\quad\quad\quad\quad\quad\quad\quad\quad
             +\frac{\tau_{i}(\tau_{i}+1)}{2}\frac{\theta^{\prime 4}}{\mathcal{H}^{2}}+3\tau_{i}\frac{\theta''}{\theta}
             -2\tau_{i}\left(\frac{\theta''}{\theta'}\right)^{2}+(\tau_{i}^{2}-1)\frac{\theta^{\prime 3}}{\mathcal{H}\theta}
             +2(\tau_{i}+1)\frac{\theta'\theta''}{\mathcal{H}}+(3+\tau_{i}-\tau_{i}^{2})\frac{\theta'''}{\theta'}\Bigg\}~,\\
& &\nonumber \tilde{\mathcal{C}}_{ij}=\frac{1}{\sqrt{|3+3\tau_{i}-\tau_{i}^{2}||3+3\tau_{j}-\tau_{j}^{2}|}}\Bigg\{
             4(\tau_{i}+\tau_{j})\mathcal{H}^{2}-2\theta^{\prime 2}+2\left[12+\tau_{i}+\tau_{j}-3\tau_{i}^{2}-3\tau_{j}^{2}
             +2\frac{c_{1}}{c_{2}}(4-\tau_{i}^{2}-\tau_{j}^{2})\right]\mathcal{H}\frac{\theta'}{\theta}\\
& &\nonumber \quad\quad\quad\quad\quad\quad\quad\quad\quad\quad\quad\quad\quad\quad\quad\quad
             -2\tau_{j}\left(\frac{\theta'}{\theta}\right)^{2}
             +(\tau_{i}^{2}+\tau_{j}^{2}-\tau_{i}-\tau_{j}-3)\frac{\theta^{\prime 4}}{\mathcal{H}^{2}}
             +2(\tau_{i}+2\tau_{j})\frac{\theta''}{\theta}-4\tau_{i}\left(\frac{\theta''}{\theta'}\right)^{2}\\
& &\nonumber \quad\quad\quad\quad\quad\quad\quad\quad\quad\quad\quad\quad\quad\quad\quad\quad
             +(\tau_{i}^{2}+\tau_{j}^{2}-2)\frac{\theta^{\prime 3}}{\mathcal{H}\theta}
             +2(\tau_{i}^{2}+\tau_{j}^{2}-2\tau_{i}-2\tau_{j}-4)\frac{\theta'\theta''}{\mathcal{H}}
             -4\tau_{j}\frac{\theta'''}{\theta'} \Bigg\}~.
\eea
This quadratic action (\ref{scalar2}) for the scalar perturbations shows that all three $\gamma_i$s represent dynamical degrees of freedom. That means the model (\ref{model}) with $c_2\neq 0$ has two more dynamical degrees of freedom than the model of GR with a minimally coupled scalar field. At the same time,
the quadratic action (\ref{scalar2}) clearly shows that $\gamma_{3}$ is a ghost mode because its kinetic term has a wrong sign.
This will cause the vacuum instability. This ghost mode in the scalar perturbations is completely brought by the PV coupling term $c_2\theta\mathcal{P}_{2}$ in the action (\ref{model}) and independent of the scale.

\subsection{Quadratic action for vector perturbations}\label{actionvector}

For vector perturbations,  direct calculation shows that the quadratic action of the model (\ref{model}) is
\bea
& &\nonumber S^{(2)}_{V}=\int d^{4}x \Bigg\{ -\frac{a^{2}}{4}B^{V}_{i}\nabla^{2}B^{V}_{i}+\frac{1}{2}c_{1}a^{2}\theta'\epsilon_{ijk}\left(\beta^{V}_{i}\beta^{V}_{j,k}-\lambda^{V}_{i}\lambda^{V}_{j,k}\right)\\
& & \quad\quad\quad\quad\quad
+c_{2}a^{2}\theta \bigg[\beta_{i}^{V\prime}\lambda_{i}^{V\prime}+\lambda_{i}^{V}\nabla^{2}\beta_{i}^{V}
+\frac{1}{2}B_{i}^{V}(\nabla^{2}\lambda^{V}_{i}-\epsilon_{ijk}\beta^{V\prime}_{j,k})
+\frac{1}{2}\Big(\frac{\theta'}{\theta}-\mathcal{H}\Big)\epsilon_{ijk}\left(\beta^{V}_{i}\beta^{V}_{j,k}-\lambda^{V}_{i}\lambda^{V}_{j,k}\right)\bigg]
\Bigg\}~.\label{vector1}
\eea
where $\nabla^{2}=\partial_{i}\partial_{i}$ is Laplacian.
Whatever the values the coefficients $c_1$ and $c_2$ take, the variable $B^{V}_{i}$ is a non-dynamical field.
It induces the constraint,
\be\label{cvector}
\nabla^{2}B^{V}_{i}+c_{2}\theta\left(\epsilon_{ijk}\beta^{V\prime}_{j,k}-\nabla^{2}\lambda^{V}_{i}\right)=0~.
\ee
This constraint equation just solves the variable $B_{i}^{V}$.
For further simplification, we use the circular polarization bases $\hat{e}^{L}_{i}$, $\hat{e}^{R}_{i}$,
so that $B_{i}^{V}$ is expanded as,
\be
B^{V}_{i}(\eta,\vec{x})=\sum_{A} \int \frac{d^{3}k}{(2\pi)^{3/2}} B_{A}(\eta,\vec{k})\hat{e}^{A}_{i}(\vec{k})e^{i\vec{k}\cdot\vec{x}}~.
\ee
The bases satisfy the relation: $\epsilon_{ijk}n^{j}\hat{e}_{k}^{A}=i \kappa_{A} \hat{e}_{i j}^{A}$,
here $A=L,R$ and $\kappa_{L}=-1$, $\kappa_{R}=1$ for the left- and right-handed polarized components respectively, $\vec{n}$  is the unit vector of $\vec{k}$.
We also expand the variables ${\lambda}^{V}_{i}$ and ${\beta}^{V}_{i}$ in the same way.
Substituting the constraint (\ref{cvector}) back into the action (\ref{vector1}),
the quadratic action for vector perturbations can be rewritten as
\bea
& &\nonumber S^{(2)}_{V}=\sum_{A} \int d\eta d^{3}ka^2\,\Bigg\{\frac{1}{2}c_{1}\kappa_{A}\theta'k\left(\beta^{*}_{A}\beta_{A}-\lambda^{*}_{A}\lambda_{A}\right)+\\
& & \quad\quad\nonumber
c_{2}\theta\bigg[{\beta^{*}_{A}}'\lambda_{A}'-\frac{1}{4}c_{2}\theta{\beta^{*}_{A}}'\beta_{A}'
-\frac{1}{2}c_{2}\kappa_{A}\theta k {\beta^{*}_{A}}'\lambda_{A}-k^{2}{\beta^{*}_{A}}\lambda_{A}
+\frac{1}{2}\kappa_{A}k\Big(\frac{\theta'}{\theta}-\mathcal{H}\Big)\left(\beta^{*}_{A}\beta_{A}-\lambda^{*}_{A}\lambda_{A}\right)
-\frac{1}{4}c_{2}\theta k^{2}\lambda^{*}_{A}\lambda_{A}\bigg]
\Bigg\}~.\label{vector1.5}
\\
& &
\eea
Again, from this action, one will see that
whether the coefficient $c_2$ vanishes or not determines whether the vector perturbations of the model (\ref{model}) contain ghost modes.

In the case where $c_2=0$, the model (\ref{model}) reduces to the NYTG model, the quadratic action (\ref{vector1.5}) reduces to the following one,
\be
S^{(2)}_{V}=\sum_{A} \int d\eta d^{3}k\, \frac{1}{2}c_{1}\kappa_{A}a^{2}\theta'k(\beta^{*}_{A}\beta_{A}-\lambda^{*}_{A}\lambda_{A})~.
\ee
Evidently, the vector perturbations $\beta_{A}^{V}$ and $\lambda_{A}^{V}$ are non-dynamical fields. There is no propagating ghost mode in this action.
In fact, without source to the vector perturbations, all of them vanish in the expanding universe. This is what happens in the NYTG model \cite{PVtele1, PVtele2}.

In the more general case with $c_2\neq0$,
the mixing terms ${\beta^{*}_{A}}'\lambda_{A}'$ and $-{\beta^{*}_{A}}'\beta_{A}'$ led by $c_2$ in the action (\ref{vector1.5}) generally indicate the existences of ghost modes.
In order to explicitly see how many dynamical degrees of freedom and how many ghost modes there are,
we redefine the following independent vector perturbation variables through $\beta^{V}_{A}$ and $\lambda^{V}_{A}$:
\be
\hat{\beta}_{A}=\frac{1}{2}c_{2}a\theta\beta_{A}-a\lambda_{A}~~,~~\hat{\lambda}_{A}=a\lambda_{A}~.
\ee
The above transformation from $(\beta_{A}, \lambda_{A})$ to $(\hat\beta_{A}, \hat\lambda_{A})$ is reversible.
Then we can rewrite the quadratic action (\ref{vector1.5}) as
\be\label{vector2}
S^{(2)}_{V}=\sum_{A} \int d\eta d^{3}k \left(
\hat{\lambda}_{A}^{*\prime}\hat{\lambda}_{A}^{\prime}-\hat{\beta}_{A}^{*\prime}\hat{\beta}_{A}^{\prime}
+\mathcal{C}_{1}\hat{\beta}_{A}^{*}\hat{\lambda}_{A}^{\prime}+\mathcal{C}_{2}\hat{\beta}_{A}^{*}\hat{\lambda}_{A}
+\mathcal{C}_{3}\hat{\lambda}_{A}^{*}\hat{\lambda}_{A}+\mathcal{C}_{4}\hat{\beta}_{A}^{*}\hat{\beta}_{A}
\right)~,
\ee
where
\bea
& &\nonumber \mathcal{C}_{1}=\kappa_{A}c_{2}\theta k-2\frac{\theta'}{\theta}~,\\
& &\nonumber \mathcal{C}_{2}=-2k^{2}-2\frac{\theta''}{\theta}-2\mathcal{H}\frac{\theta'}{\theta}+\kappa_{A} c_{2}k(2\theta'+\mathcal{H}\theta)
    +\frac{4\kappa_{A}k}{(c_{2}\theta)^{2}}\left[(c_{1}+c_{2})\theta'-c_{2}\mathcal{H}\theta\right]~,\\
& &\nonumber \mathcal{C}_{3}=-2k^{2}+2\mathcal{H}^{2}-\frac{1}{2}\theta^{\prime 2}-\frac{\theta^{\prime 2}}{\theta^{2}}
    -\frac{1}{4}c_{2}^{2}\theta^{2}k^{2}+\frac{\kappa_{A}k}{2}\left(3c_{2}\mathcal{H}\theta+2c_{2}\theta'-c_{1}\theta'\right)
    +\frac{2\kappa_{A}k}{(c_{2}\theta)^{2}}\left[(c_{1}+c_{2})\theta'-c_{2}\mathcal{H}\theta\right]~,\\
& &\nonumber \mathcal{C}_{4}=-2\mathcal{H}^{2}+\frac{1}{2}\theta^{\prime 2}-\frac{\theta''}{\theta}-2\mathcal{H}\frac{\theta'}{\theta}
    +\frac{2\kappa_{A}k}{(c_{2}\theta)^{2}}\left[(c_{1}+c_{2})\theta'-c_{2}\mathcal{H}\theta\right]~.
\eea
The quadratic action (\ref{vector2}) shows that all the four components of vector perturbations, $\hat\beta_{A}$ and $\hat\lambda_{A}$ with $A=L ~{\rm and}~R$, are dynamical modes.
It also clearly shows that both components of $\hat{\beta}_{A}$ are ghost modes because their kinetic terms have wrong signs.
Again, this will cause the vacuum instability and such instability in the vector perturbations is completely brought by the PV coupling term $c_2\theta\mathcal{P}_{2}$ in the action (\ref{model}) and is scale independent.

As a comparison, the model of GR with a scalar field or the NYTG model have no propagating dynamical degrees of freedom in the vector perturbations. Combining with the analysis on the scalar perturbations in the above subsection, the model (\ref{model}) with $c_2\neq 0$ brings totally six more dynamical degrees of freedom than the model of GR with a minimally coupled scalar field.
This confirms the judgement in Sec.~\ref{Model}. Furthermore, three of these extra dynamical degrees of freedom represent ghost modes.

\subsection{Quadratic action for tensor perturbations}

Finally, we have the quadratic action for tensor perturbations in the model (\ref{model}),
\be\label{tensor1}
  S^{(2)}_{T}=\int d^{4}x \ \frac{a^{2}}{8} \left[{h^{T}_{ij}}'{h^{T}_{ij}}'-h^{T}_{ij}\nabla^{2}h^{T}_{ij}
  -(c_{1}\theta'-3c_{2}\mathcal{H}\theta) \epsilon_{ijk}h^{T}_{il}h^{T}_{lj,k}\right]~.
\ee
In terms of the circular polarization bases $\hat{e}^{L}_{ij}$ and $\hat{e}^{R}_{ij}$, where $n^{l} \epsilon_{lik}\hat{e}_{j k}^{A}=i \kappa_{A} \hat{e}_{i j}^{A}$ with $A=L,R$ and $\kappa_{L}=-1$, $\kappa_{R}=1$, the tensor perturbations are expanded as
\be
h_{i j}^{T}(\eta, \vec{x})=\sum_{A} \int \frac{d^{3} k}{(2 \pi)^{3 / 2}}\, h_{A}(\eta, \vec{k})\, \hat{e}_{i j}^{A}(\vec{k})\, e^{-i\vec{k}\cdot\vec{x}}~.
\ee
The final form of the quadratic action for tensor perturbations is
\be\label{tensor2}
S^{(2)}_{T}=\sum_{A} \int d \eta d^{3} k~ \frac{a^{2}}{4}\left[h_{A}^{*\prime} h_{A}^{\prime}
-k^{2}\left(1+\kappa_{A}\frac{c_{1}\theta'-3c_{2}\mathcal{H}\theta}{k}\right) h_{A}^{*} h_{A}\right]~.
\ee
First, we see that the tensor perturbations are ghost free.
Second, the modified dispersion relation
$\omega_{A}^{2}=k^{2}\left[1+\kappa_{A}(c_{1}\theta'-3c_{2}\mathcal{H}\theta)/k\right]$ is helicity dependent.
Consider small coupling and slow evolution of $\theta$, one can find that GWs with different helicities will
have different phase velocities $v_{p}^{A}=\omega_{A} / k \simeq 1+\kappa_{A} (c_{1}\theta'-3c_{2}\mathcal{H}\theta)/(2 k)$, i.e., the velocity birefringence phenomenon.
This is the explicit signal of parity violation in this model.
We can also see that the phase velocity difference become important only at the region of small $k$ (large scales), so this is an infrared effect.
Besides the velocity birefringence, the amplitudes for both components of GWs are the same.
When $c_2=0$, these results go back to those obtained in the NYTG model \cite{PVtele1, PVtele2}.

\section{Perturbations around the Minkowski background}\label{Minkowski}

To further explore the severity of the ghost instability in the model (\ref{model}), we study the perturbations around the Minkowski background in this section.
We still take the Weitzenb\"{o}ck gauge $\omega^{A}_{~B\mu}=0$,
and still use Eq.~(\ref{tetradperturbation}) to parameterize the perturbations of the tetrad, but the scale factor $a=1$ in the Minkowski background.
In addition, the scalar field $\theta$ is decomposed as $\theta=\theta_0+\delta\theta$,
where the constant $\theta_0$ is the lowest point of potential $V(\theta)$, satisfying $V(\theta_{0})=0$ and $V_{\theta}(\theta_{0})=0$.
Note that $\delta\theta$ is gauge invariant in this case, so we can no longer take the unitary gauge in this section.
Instead we take the conformal Newton gauge: $B=0,~\alpha=0,~\alpha_{i}^{V}=0$,
and we choose $A$, $\psi$, $\delta\theta$, $\beta$, $\lambda$, $B_{i}^{V}$, $\beta_{i}^{V}$, $\lambda_{i}^{V}$ and $h^{T}_{ij}$ as independent variables.
Since the purpose of this section is to explore the severity of the ghost instability in the model (\ref{model}), for simplicity,
we only consider the case of $S_m=0$, $c_1\neq0$, $c_2\neq0$ and $\theta_{0}\neq0$ in the following.

For scalar perturbations,  direct calculation shows that the quadratic action of the model (\ref{model}) is
\bea\label{Mscalar1}
S^{(2)}_{S}=\int dt d^{3}k~\bigg\{ \frac{1}{2}\Big[\delta{\theta'}^{2}-(k^{2}+m_{\theta}^{2})\delta\theta^{2}\Big]
-3{\psi'}^{ 2}+k^{2}\psi^{2}-2k^{2}A\psi
+c_{2}\theta_{0}\Big[k^{2}(\beta'-A-\psi)\lambda'-k^{4}\beta\lambda\Big]\bigg\}~,
\eea
where $m_{\theta}^{2}=d^{2}V(\theta)/d\theta^{2}|_{\theta=\theta_{0}}$.
We have changed to the momentum space in terms of Fourier transformations and
we also simply mark $A^{*}\psi$ as $A\psi$, $\psi^{*}\psi$ as $\psi^{2}$, and so on.
The variable $A$ is non-dynamical field and it induces the constraint
\be\label{co1}
2\psi+c_{2}\theta_{0}\lambda'=0~.
\ee
Substituting the constraint (\ref{co1}) back into the action (\ref{Mscalar1}),
the quadratic action for scalar perturbations can be rewritten as
\bea\label{Mscalar2}
S^{(2)}_{S}=\int dt d^{3}k~\bigg\{ \frac{1}{2}\Big[\delta{\theta'}^{2}-(k^{2}+m_{\theta}^{2})\delta\theta^{2}\Big]
+\frac{1}{4}\Big[3c_{2}^{2}\theta_{0}^{2}(-{\lambda''}^{2}+k^{2}{\lambda'}^{2})-4c_{2}\theta_{0}k^{2}(\beta\lambda''+k^{2}\beta\lambda)\Big]\bigg\}~.
\eea
The action (\ref{Mscalar2}) is equivalent to the following action,
\bea\label{Mscalar3}
S^{(2)}_{S}=\int dt d^{3}k~\bigg\{ \frac{1}{2}\Big[\delta{\theta'}^{2}-(k^{2}+m_{\theta}^{2})\delta\theta^{2}\Big]
+\frac{1}{4}\Big[c_{2}\theta_{0}\phi\lambda''+\frac{1}{12}\phi^{2}+\frac{2}{3}k^{2}\beta\phi+\frac{4}{3}k^{4}\beta^{2}
+3c_{2}^{2}\theta_{0}^{2}k^{2}{\lambda'}^{2}-4c_{2}\theta_{0}k^{4}\beta\lambda\Big]\bigg\}~.
\eea
The variation of the action (\ref{Mscalar3}) on $\phi$ leads to constraint
\be\label{co2}
\phi=-2(2k^{2}\beta+3c_{2}\theta_{0}\lambda'')~.
\ee
It can be verified that the action (\ref{Mscalar2}) can be obtained by substituting the constraint (\ref{co2})
back into the action (\ref{Mscalar3}), so the action (\ref{Mscalar2}) and the action (\ref{Mscalar3}) are equivalent.
On the other hand, the variation of the action (\ref{Mscalar3})  on $\beta$ leads to the constraint
\be\label{co3}
\beta=\frac{3}{2}c_{2}\theta_{0}\lambda-\frac{1}{4k^{2}}\phi~.
\ee
Substituting the constraint (\ref{co3}) back into the action (\ref{Mscalar3}),
the quadratic action for scalar perturbations can be rewritten as
\bea\label{Mscalar4}
S^{(2)}_{S}=\int dt d^{3}k~\bigg\{ \frac{1}{2}\Big[\delta{\theta'}^{2}-(k^{2}+m_{\theta}^{2})\delta\theta^{2}\Big]
+\frac{c_{2}\theta_{0}}{4}\Big[3c_{2}\theta_{0}k^{2}{\lambda'}^{2}-\phi'\lambda'+k^{2}\lambda\phi-3c_{2}\theta_{0}k^{4}\lambda^{2}\Big]\bigg\}~.
\eea
In order to further simplify the quadratic action,
we define new perturbation variables $q_1, q_2$ in terms of the old variables $\lambda, \phi$:
\bea\label{newq}
& &\nonumber
q_{1}=2^{-\frac{3}{2}}\left(1+9c_{2}^{2}\theta_{0}^{2}k^{4}\right)^{-\frac{1}{4}}
\left[\phi-\left(3c_{2}\theta_{0}k^{2}+\sqrt{1+9c_{2}^{2}\theta_{0}^{2}k^{4}}\right)\lambda\right]~,
\\
& &
q_{2}=2^{-\frac{3}{2}}\left(1+9c_{2}^{2}\theta_{0}^{2}k^{4}\right)^{-\frac{1}{4}}
\left[\phi-\left(3c_{2}\theta_{0}k^{2}-\sqrt{1+9c_{2}^{2}\theta_{0}^{2}k^{4}}\right)\lambda\right]~.
\eea
Then we can express the action (\ref{Mscalar4}) in terms of the new perturbation variables as,
\bea\label{Mscalar5}
S^{(2)}_{S}=\int dt d^{3}k~\bigg\{ \frac{1}{2}\Big[\delta{\theta'}^{2}-(k^{2}+m_{\theta}^{2})\delta\theta^{2}\Big]
+c_{2}\theta_{0}\Big[ \frac{1}{2}({q_{1}'}^{2}-k^{2}q_{1}^{2})-\frac{1}{2}({q_{2}'}^{2}-k^{2}q_{2}^{2})\Big]\bigg\}~.
\eea
The quadratic action (\ref{Mscalar5}) for the scalar perturbations shows that $\delta\theta, q_{1}, q_{2}$
are all dynamical degrees of freedom and $q_{2}$ is a ghost mode because its kinetic term has a wrong sign.
Such a result is consistent with the results in Sec.~\ref{actionscalar}.
Furthermore, the action (\ref{Mscalar5}) shows that the mass of the ghost mode $q_2$ is zero.
This means that no matter how low the energy scale of the effective field theory is, the ghost mode $q_{2}$ can always be excited.
Therefore, even from the point of view of effective field theory,  the model (\ref{model}) still suffers from the difficulty of ghost instability.

For vector perturbations,
its quadratic action can be obtained through the same process as in Sec.~\ref{actionvector},
and the result is still the action (\ref{vector2}),
but here $\mathcal{C}_{1}=\kappa_{A}c_{2}\theta_{0} k$, $\mathcal{C}_{2}=-2k^{2}$,
$\mathcal{C}_{3}=-(2+c_{2}^{2}\theta_{0}^{2}/4)k^{2}$ and $\mathcal{C}_{4}=0$.
The action (\ref{vector2}) shows that $\hat\beta_{A}$ and $\hat\lambda_{A}$ are all dynamical modes but $\hat{\beta}_{A}$ are ghost modes.
The fact that the mass term coefficient $\mathcal{C}_{4}$ of the ghost modes $\hat{\beta}_{A}$ vanish
implies that the ghost modes $\hat{\beta}_{A}$ can always be excited no matter how low the energy scale of the effective field theory is.
Such results again illustrate that even from the point of view of effective field theory,
the model (\ref{model}) suffers from the difficulty of ghost instability.

For tensor perturbations, its quadratic action is exactly the same as that of GR, so we won't discuss it anymore.

\section{Conclusion}\label{conclusion}

We studied in this paper a parity violating gravity model within the framework of teleparallel gravity. This model modifies the gravity by the couplings of a scalar field to the scalar invariants which are parity-odd and quadratic in the torsion tensor. Totally there are two such type independent invariants, and one of them is the Nieh-Yan density. Through investigations on its cosmological perturbations, we found that in general this model has six more dynamical degrees of freedom compared with model of GR with a minimally coupled scalar field. Among these extra dynamical degrees of freedom, two of them are scalar perturbations and the rest four are vector perturbations. Furthermore, half of them are ghost modes and cause instabilities at all scales.  However, in the special case only the coupling to the Nieh-Yan density exists, this model is ghost free and reduces to the NYTG model.
Finally, by investigating the properties of the ghost modes in Minkowski spacetime,
we find that it is almost impossible to avoid ghost instability in the model (\ref{model}) by adjusting the energy scale of the effective field theory.

{\it Acknowledgement}: This work is supported in part by NSFC under Grant No. 12075231 and 12047502, and by National Key R\&D Program of China Grant No. 2021YFC2203100.

{}


\end{document}